          [2001/04/25 1.1 (PWD)]
\documentclass[a4paper,twocolumn]{spaceDebrisC} 
\usepackage{times}
\usepackage[numbers]{natbib}
\usepackage{graphicx}

\title{Design of pulsed waveforms for space debris detection with ATLAS}
\author[1,2]{João Pandeirada}
\author[1,3]{Miguel Bergano}
\author[4]{Paulo Marques}
\author[1]{Domingos Barbosa}
\author[5]{José Freitas}
\author[1]{Bruno Coelho}
\author[1]{Valério Ribeiro}

\affil[1]{Instituto de Telecomunicações, Aveiro, Portugal; joao.pandeirada@av.it.pt (J.P); jbergano@av.it.pt (M.B); dbarbosa@av.it.pt (D.B); brunodfcoelho@av.it.pt (B.C);  valerio.ribeiro@av.it.pt (V.R)}
\affil[2]{DETI - Universidade de Aveiro, Aveiro, Portugal}
\affil[3]{ESTGA - Universidade de Aveiro, Aveiro, Portugal}
\affil[4]{Instituto de Telecomunicações / ISEL-IPL, Lisboa, Portugal; pmarques@isel.pt (P.M)}
\affil[5]{Ministério da Defesa Nacional, Portugal; jose.freitas@defesa.pt (J.F)}

\begin{document}


\maketitle

\begin{abstract}
ATLAS is the first Portuguese radar system that aims to detect space debris. The article introduces the system and provides a brief description of its capabilities. The system is capable of synthesizing arbitrary amplitude modulated pulse shapes with a resolution of 10 ns. Given that degree of freedom we decided to test an amplitude modulated chirp signal developed by us and a nested barker code. These waveforms are explained as well as their advantages and drawbacks for space debris detection. An experimental setup was developed to test the system receiver and waveforms are processed by digital matched filtering. The experiments test the system using different waveform shapes and noise levels. Experimental results are in agreement with simulation and show that the chirp signal is more resilient to Doppler shifts, has higher range resolution and lower peak-to-sidelobe ratio in comparison with the nested barker code. Future work in order to increase detection capabilities is discussed at the end.
\end{abstract}

\section{Introduction}

An element of utmost importance in space programs is the safety of space vehicles both in-orbit and during the launching process. In-orbit space assets are the foundations of many space-based services such as communications, navigation, timing and earth observation which ensure the progress and safety of the economies and societies worldwide. Due to the growing number of space objects and increased complexity of the orbital environment, the risk of collisions in space increases as well the byproduct of those collisions, space debris \cite{esa}.

The most recent sensor for space debris detection in Portugal is ATLAS \cite{jpandeirada} and consists of a monostatic pulse radar located  in the center of Portugal - Pampilhosa da Serra. ATLAS is included in the list of EUSST radar assets network that shall communicate with the PT SST Network Operating Centre (NoC) in Azores. The system uses state-of-the-art solid state GaN technology to deliver a peak output power of 5 kW and presents an architecture that is highly modular with signal processing fully on the digital domain. The receiver is coherent with the emitter and detection and processing is fully in the digital domain with a bandwidth of 50 MHz and capacity of detecting Doppler velocities up to 10.79 km/s.

\begin{figure}[!b]
\centering
\includegraphics[width=0.8\linewidth]{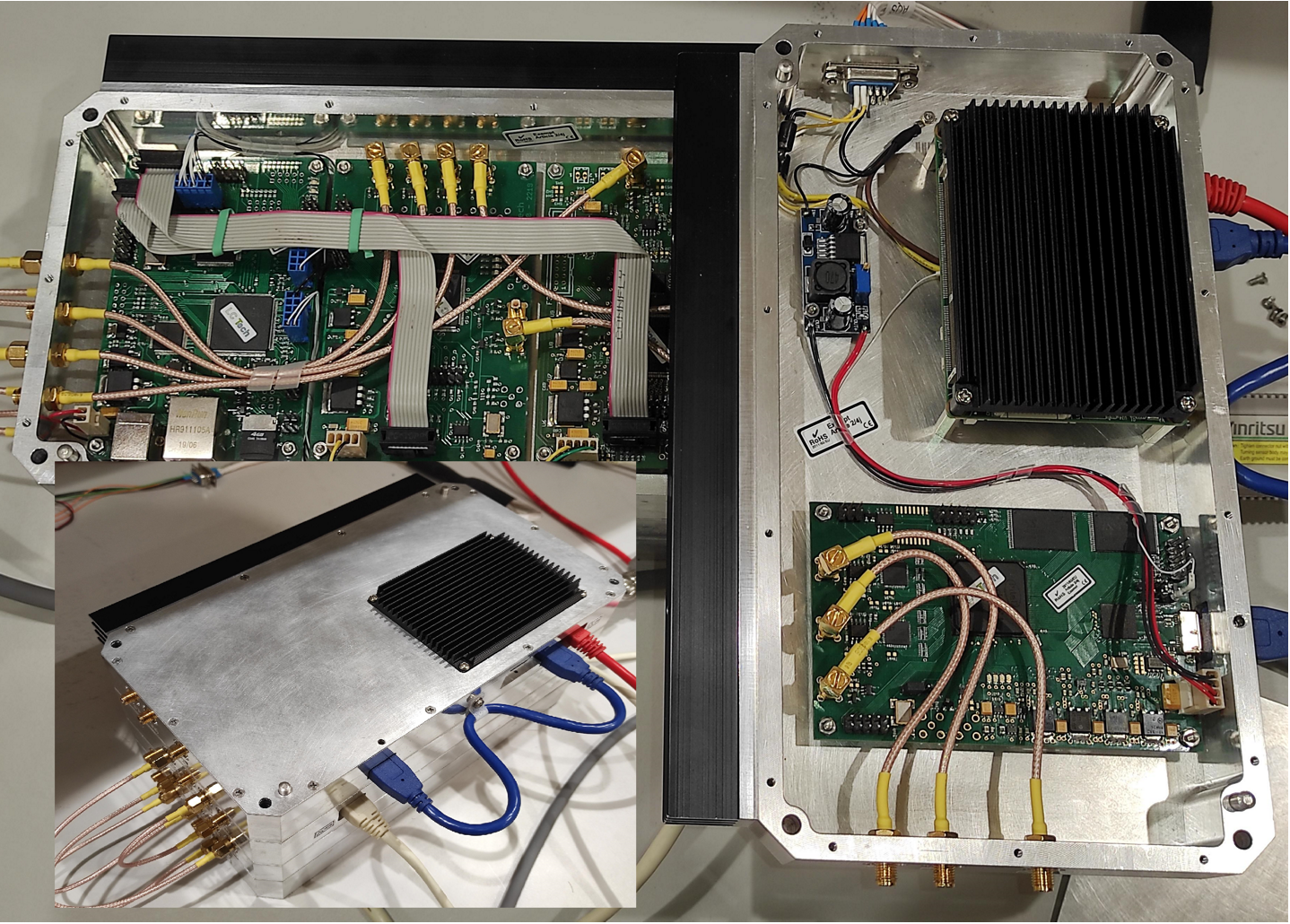}
\caption{ATLAS back-end components.}
\label{fig:backend}
\end{figure}

ATLAS establishes a platform for fast and easy development, research and innovation due to the high modularity and usage of Commercial-Off-The-Shelf technologies and Open Systems (Fig.\ref{fig:backend}). The system is capable of synthesizing arbitrary amplitude modulated pulse shapes with a resolution of 10 ns. Given that degree of freedom, we decided to test an amplitude modulated chirp signal developed by us and a nested barker code. The article explains these waveforms as well as their advantages and drawbacks for space debris detection. An experimental setup was developed to test the system receiver and a digital signal processor based on matched filtering. The experiments test the system using different waveform shapes and noise levels.

\section{Waveforms}
\subsection{AM Chirp}
Linear frequency modulation, commonly known as chirp, is a widely popular waveform used in radar systems due to its highly compressed autocorrelation and Doppler resilience \cite{chirplevanon}. Since ATLAS only has amplitude modulation at the moment, in order to implement LFM signals we developed an AM Chirp. The AM Chirp consists of designing a chirp signal at a much lower frequency and using it for modulating the carrier in amplitude \cite{jpandeirada}. The AM Chirp signal, y(t) is given by:

\begin{equation}
    y(t) = A(t)\cos(2\pi f_c t),
\end{equation}

\begin{equation}
    A(t) = \alpha + \beta \cos[2\pi(ct+f_0)t],~ c = \frac{f_1 - f_0}{T}
\end{equation}

where $f_c$ is the carrier frequency, $\alpha$ and $\beta$ define the modulation parameters, $f_0$ and $f_1$ are the starting and final frequencies and $T$ is the pulse duration.

\begin{figure}[!b]
\centering
\includegraphics[width=0.9\linewidth]{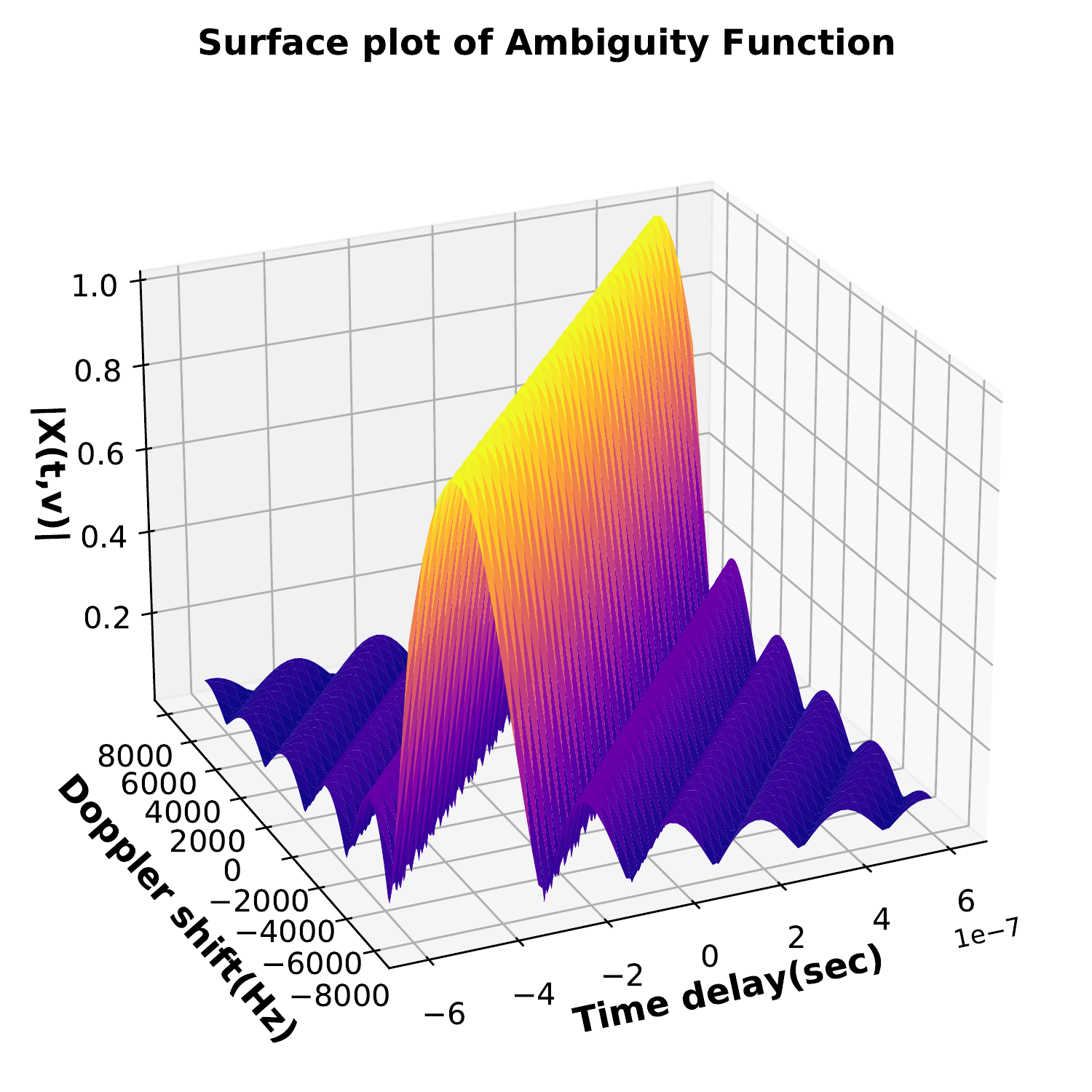}
\caption{AF of AM Chirp with parameters: $\alpha = 0.5$; $\beta = 0.5$;
$f_0 = -5 $ MHz; $f_1 = 5 $ MHz; $T = 640 \mu s$; SNR = 20 dB.}
\label{fig:amchirp}
\end{figure}

Fig.\ref{fig:amchirp} depicts the simulated ambiguity function (AF) of an AM chirp at 20 dB of signal to noise ratio (SNR) that will be used in ATLAS in the near future. This waveform presents an high Doppler tolerance as well as an high auto correlation compression. By inspection of the AF it is clear that there is range-Doppler coupling, so Doppler mismatches do not change pulse shape/amplitude significantly, but they do appear to shift the pulse in time.

The developed AM chirp presents a range resolution of 60 m at 5.56 GHz (ATLAS operating frequency).

\subsection{Nested Barker Code}
A Barker code is a finite sequence of $N$ values $a_j$ that can be +1 or -1 and that match the following condition:

\begin{equation}
    |\sum_{j=1}^{N-v} a_j a_{j+v}| \leq 1
\end{equation}

where $1 \leq v \le N$ \cite{barker}. This guarantee that the peak-to-peak sidelobe ratio of the autocorrelation is $M$ , where $M$ is the size of the Barker code \cite{jpandeirada}. Since the longer known barker code has only 13 elements, in order to create longer codes with the same autocorrelation properties it is possible to create nested barker codes. Nested barker codes consist in doing the Kronecker product of two barker codes \cite{nestedbarker}.

\begin{figure}[!b]
\centering
\includegraphics[width=0.9\linewidth]{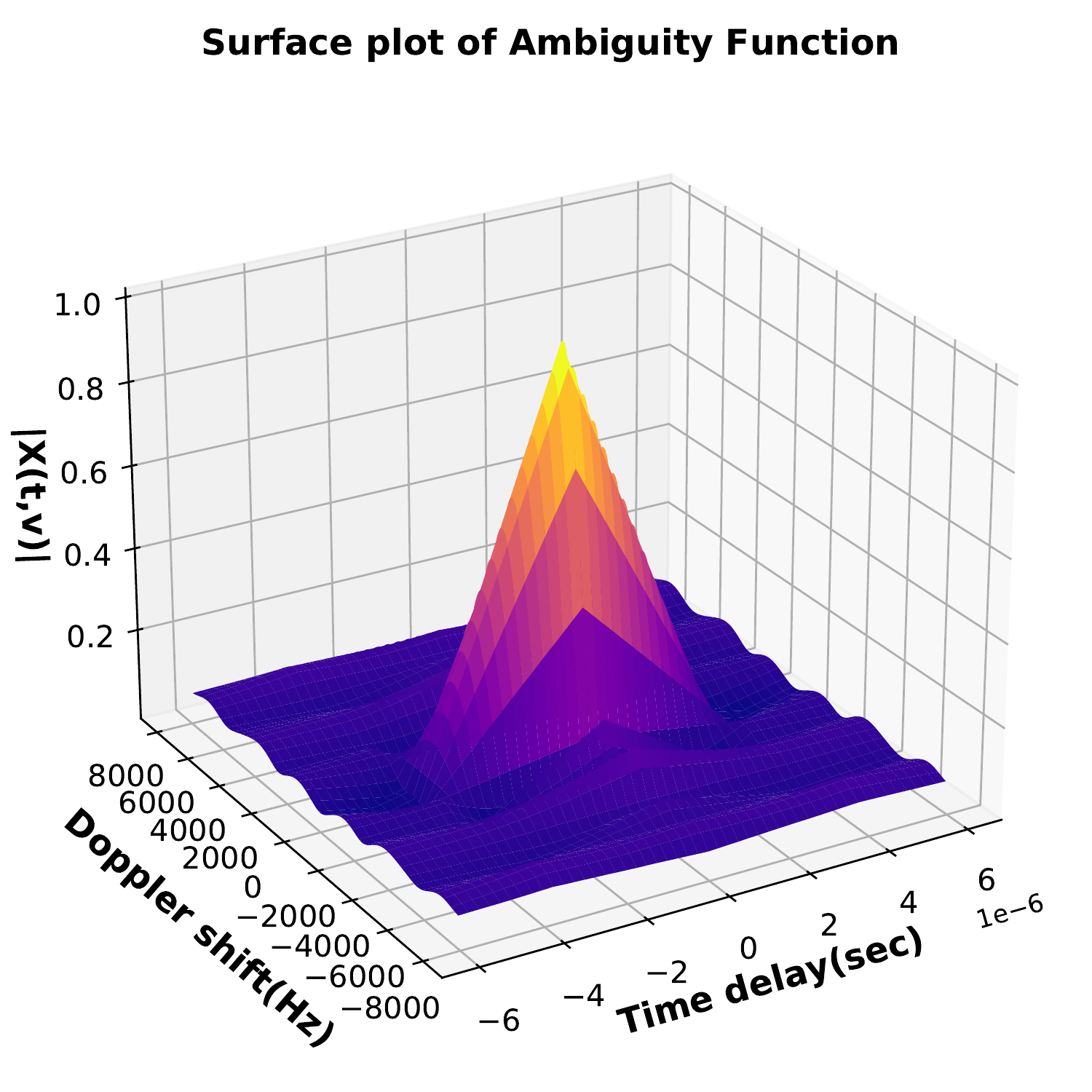}
\caption{AF of a 13x13 element nested barker code with a pulse duration of 640 $\mu$s and SNR = 20 dB.}
\label{fig:barker}
\end{figure}

Fig.\ref{fig:barker} depicts the AF of a 13 by 13 element nested barker code that can be implement in ATLAS. The barker code offers low tolerance to Doppler shifts as it can be seen by the sudden decrease on the AF amplitude along the frequency axis.

The nested barker code presents a range resolution of 1.1 km at 5.56 GHz (ATLAS operating frequency).

\begin{figure*}[!t]
\centering
\includegraphics[width=0.8\linewidth]{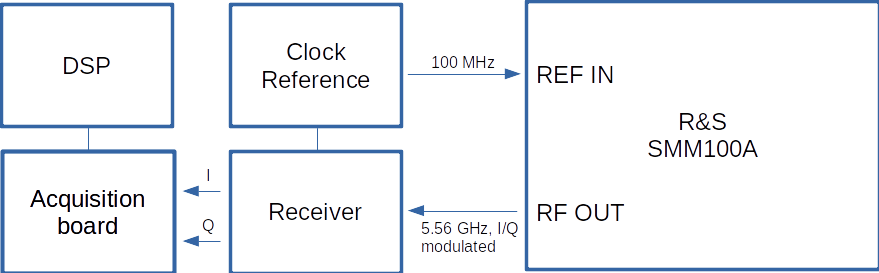}
\caption{Experimental Set-Up}
\label{fig:setup}
\end{figure*}

\section{Experimental Setup}
In order to validate the waveforms with the system hardware, an experimental setup was developed (Fig.\ref{fig:setup}). The R\&S SMM100A vector signal generator is responsible for emulating the echo received by the radar. The waveforms are loaded to the equipment and it outputs a arbitrarily modulated 5.56 GHz RF signal with a desired power. This signals enters ATLAS receiver where it is demodulated and filtered. Finally, it is digitized by ATLAS acquisition board. The digitized echo passes through a digital matched filter implemented in a DSP to obtain the final pulse compressed signal.

In order to maintain the receiver and the signal generator coherent, a 100 MHz reference signal from the ATLAS clock reference system is plugged to the external reference input of the R\&S SMM100A.

\section{Results and Discussion}
\begin{figure}[!b]
\centering
\includegraphics[width=\linewidth]{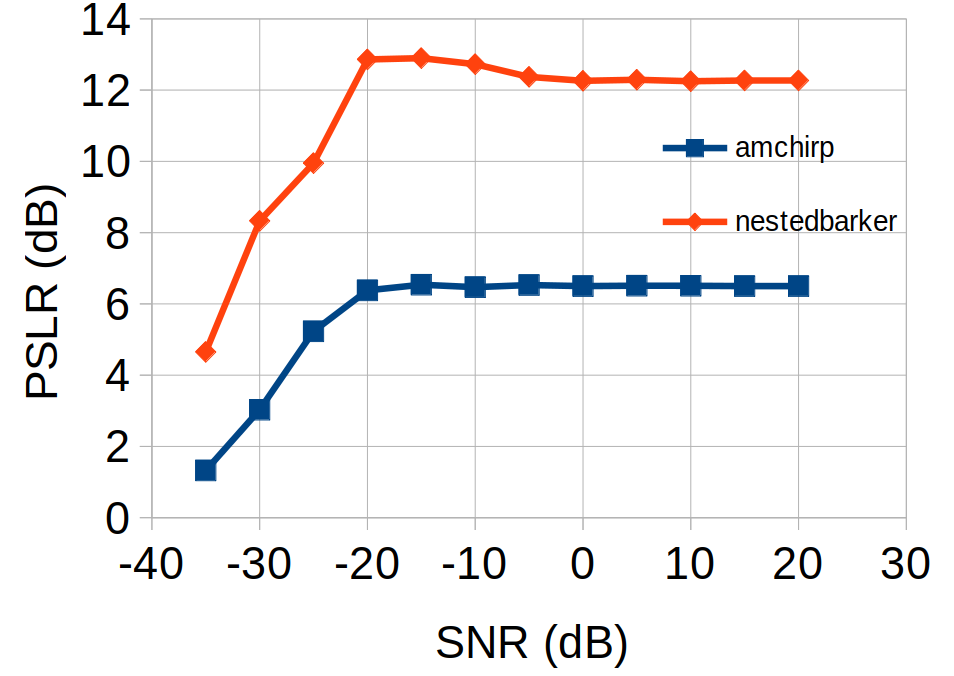}
\caption{PSLR for AM Chirp and Nested Barker Code.}
\label{fig:pslr_sim}
\end{figure}

The peak-to-sidelobe ratio (PSLR) is a measurement on how small are the sidelobes of a waveform in comparison to the main lobe. This metric is used to evaluate the detection performance of a waveform. Fig.\ref{fig:pslr_sim} shows simulation results for the PSLR obtained by the previously presented waveforms at different SNR.

\begin{figure}[!b]
\centering
\includegraphics[width=\linewidth]{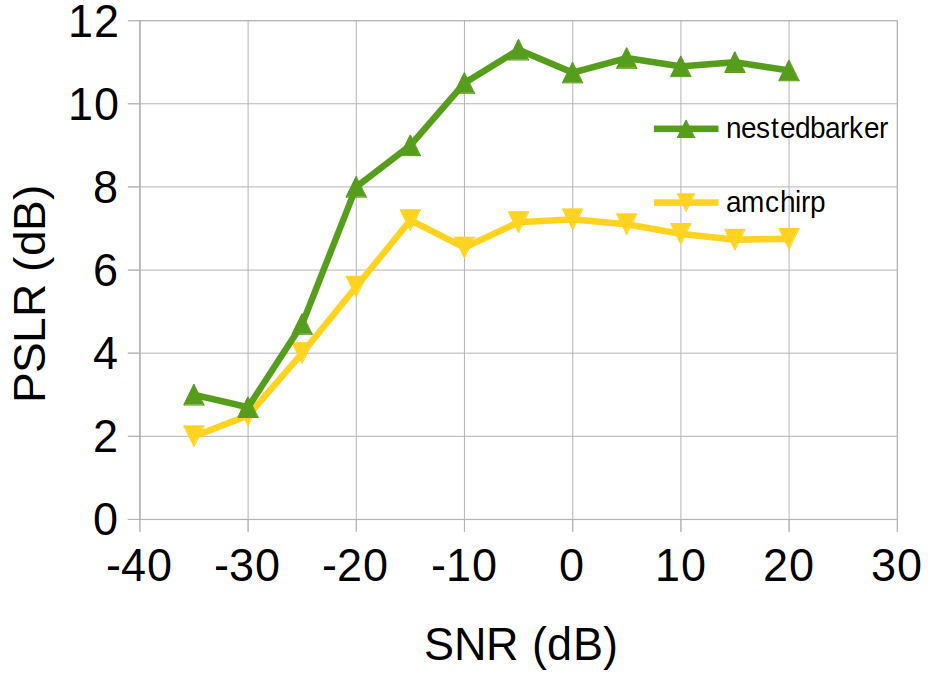}
\caption{Experimental PSLR for AM Chirp and Nested Barker Code.}
\label{fig:pslr_exp}
\end{figure}

After acquiring both the AM Chirp and the nested barker code for different SNR values with the setup depicted in Fig.\ref{fig:setup}, the experimental PSLR curves were obtained and are displayed in Fig.\ref{fig:pslr_exp}. The experimental results for the AM chirp are in accordance to the simulated ones. The experimental curve for the barker code follows the same trend as the simulated one but has a difference of around 2 dB.

The nested barker code presents a much higher PSLR than the AM Chirp, which makes it more reliable for peak detection, specially at lower SNR. Keep in mind that the PSLR was obtained for a zero Doppler shift response. If there is Doppler mismatch, the nested barker code performance will drop drastically, as it can be seen in Fig.\ref{fig:barker}. On the other hand, the Doppler resilience of the chirp comes at the cost of lower PSLR.

\section{Conclusions and Future Work}
The AM chirp presents a resolution of 60 m while the barker code has a range resolution of 1.1 km. The chirp is much more Doppler resilient than the barker code but the latter has an higher PSLR which increases detectability.

Future work consists in developing strategies to deal with the limitation of each waveform.
Using an window function will increase the PSLR of the AM Chirp at the cost of a wider main lobe which decreases range resolution. Increasing the Doppler resilience of the barker code is possible by having a bank of matched filters tuned for different Doppler shifts. These and other techniques will be investigated in the future.

\section*{Acknowledgments}
This work has been funded by the European Commission Horizon 2020 Programme under grant agreement 2-3SST2018-20.

\end{document}